\documentclass[a4paper,showkeys,floatfix,aps,pre,reprint,groupedaddress]{revtex4-1}
\usepackage{graphicx}
\usepackage{amsmath, amssymb, amsfonts}
\usepackage{color}
\usepackage{subfig}
\begin{document}

\title{Crossover behaviors in one and two dimensional heterogeneous load sharing  fiber bundle models}



\author{Soumyajyoti Biswas}
\email[]{soumyajyoti.biswas@saha.ac.in}
\author{Bikas K. Chakrabarti}
\email[]{bikask.chakrabarti@saha.ac.in}

%\thanks{}
\affiliation{
Theoretical Condensed Matter Physics Division, Saha Institute of Nuclear 
Physics, 1/AF Bidhannagar, Kolkata-700064, India.\\
}

\date{\today}

\begin{abstract}
\noindent   We study the effect of heterogeneous load sharing in the fiber bundle models of 
fracture. The system is divided into two groups of fibers (fraction $p$ and $1-p$) in 
which one group follow the completely local load sharing mechanism and the 
other group follow global load sharing mechanism. Patches of local  disorders (weakness)
in the loading plate can cause such a situation in the system. We find that in 2d a finite crossover 
(between global and local load sharing behaviours)
point comes up at a finite value of the disorder concentration (near $p_c\sim 0.53$), which is slightly 
below the site percolation threshold.
We numerically determine the phase diagrams (in 1d and 2d) and identify the 
critical behavior below $p_c$ with the mean field behavior (completely global load 
sharing) for both dimensions. This crossover can occur due to geometrical percolation of disorders in the 
loading plate. We also show how the critical point depends on the loading history, which is 
identified as a special property of local load sharing.

\end{abstract} %end of abstract
\maketitle
\section{Introduction}
\noindent The fiber bundle model serves as a simple model for failure due to fracture. 
Originally starting with the textile industry \cite{first}, some of its characteristic features mimic 
 the failure processes in many disordered materials \cite{dani,coleman,books,rmp1}. Also, it can  have direct similarities
with fiber reinforced composites (FRC), which are very useful materials for aerospace industries \cite{kun}.
Due to its importance in theoretical modelling of disordered materials in general and FRCs 
in particular, fiber bundle model have been studied in great details \cite{newman,roux,pradhan1,hemmer1,hansen1,sornette,zapperi,pradhan2,pradhan3,pradhan4,rmp2}. 

The basic model is about a bunch of fibers, having different failure thresholds and hanging from
a rigid plate. The fibers are also attached to a plate at the bottom, from which a load is hanging. 
If the threshold distribution starts continuously from zero (in most cases they do), then as soon
as any load is applied, some of the weak fibers will break. The load carried by those fibers will now
be redistributed among the other intact fibers and this may lead to failure of some more fibers and 
so on. This avalanche can stop due to one of the following two reasons: either all the fibers may break
and the system undergoes complete failure, or the remaining fibers can withstand the load and the system 
is only partially damaged. The transition between partial and complete failure have been studied
using the standard tools of phase transitions (diverging time scales, scale free behavior of avalanche
size distribution, susceptibility and so on).  

Depending upon the requirements and interests of study, the fiber bundle model can differ in some 
key aspects. The threshold distribution of the fibers can vary (uniform distribution, Gaussian, 
Weibull, Gumbel etc.). Although the failure point sensitively depends upon the details of these
distributions, the exponents remain universal (here we use uniform distribution). Another key aspect in which the models can differ
is the load sharing mechanisms. When a fiber breaks, its load is to be shared by  the remaining intact 
fibers. But there can be many ways in which this can be done. Two extreme cases are: Global Load
Sharing (GLS) and Local Load Sharing (LLS). In GLS, the load of the broken fiber is distributed equally
among the rest of the fibers and in LLS it is distributed only among the nearest surviving neighbor(s).
In general, the elastic properties of the bottom plate (or that of the matrix material in FRCs) are
responsible for the load sharing mechanisms. The two extreme cases correspond to perfectly rigid and 
perfectly soft bottom plates.  The real situation is, of course, somewhere between these two. Hence, there 
have been efforts to interpolate between the two mechanisms \cite{inter1,inter2,inter3,inter4}. 
Finally, another aspect that we want to mention 
here is the different loading processes. There are several ways in which the load can be increased. 
One way is to apply a load and then wait for the avalanche to stop. Once the avalanche stops, increase
the load upto the point where the weakest surviving fiber breaks. This will start another avalanche and
so on. Another way is to increase the load by a constant amount every time an avalanche stops. And finally, 
one can also directly apply the required load in the intact system. In GLS models, the loading mechanisms
do not change the critical point. In some situations, this can be important, as we shall discuss.

In the different models, the load sharing mechanisms followed by all the fibers were same (global, local 
or some other intermediate rule). However, the governing factors of the load sharing mechanisms, like
the elastic properties of the bottom plate, can be modified due to local disorders, i.e., the elastic
properties of the bottom plate may not be homogeneous. The heterogeneity introduced by local disorders
can make the load sharing mechanisms heterogeneous as well. In this study we investigate the effect of
such heterogeneity. We divide the fiber into two groups, where one group follow GLS and the other follow LLS.
The competing effects of these two mechanisms give rise to a cross-over behavior at a threshold value
of the fraction. We determine the phase diagram and the critical behavior numerically. Also, we show that
the process of loading changes the critical point in this case. We argue that spatial correlations in the
presence of local load sharing gives rise to this change in critical point.   

\section{Heterogeneous load sharing}
\noindent As pointed out before, the load sharing process in the system is a manifestation of the elastic
properties of the base plate. Hence, local disorders in that material may lead to heterogeneous load
sharing processes. To see some effects of that we divide the system into two parts, a fraction $p$ of the
fibers follow completely local load sharing process and the rest $1-p$ fraction follow global load sharing 
process. In increasing the load we have followed the direct loading mechanism, i.e., we apply a load directly 
in the intact fiber. Comparisons with other loading mechanisms are discussed in the later part. 

We study this model both in 1d and 2d (square lattice) geometry numerically. Throughout the paper we have used uniform distribution (in the range [$0:1$]) of the breaking threshold for the fibers. Since this is also an interpolation
mechanism ($p=0$ corresponds to GLS and $p=1$ corresponds to LLS), we compare the results with that of ref \cite{inter2},
where a fraction $g$ of the load of the broken fiber is shared locally and the rest is distributed globally.
Below we present the simulation results in one and two dimensions for this model.

\subsection{Results in one dimension}
\noindent In this case the fibers are arranged in a linear lattice. After applying a load on the intact fiber,
we scan the lattice once and the fibers having threshold below the applied load are broken. This would create
patches of broken fibers. Now, among them the fraction of load carried by the local load sharing fibers, are 
distributed to the nearest surviving neighbors (in this way, the load of an isolated local load sharing fiber will go to the nearest surviving neighbors) and the rest is distributed globally among the remaining fibers.
This may lead to further failures and so on.
%%%%%%%%%%%%%%%%%%%%%%%%%%%%%%%%%%%%%%%%%%%%%%%%%%%%%%%%%%%%%%%%%%%%%%%%%%%%%%%%%%%%%%%%%%%%%%%%%%%%%%%%%%%%%%%%%%%%%%%%%%%%%%%%%%%%%%%%%
\begin{figure}[tb]
\centering \includegraphics[width=8cm]{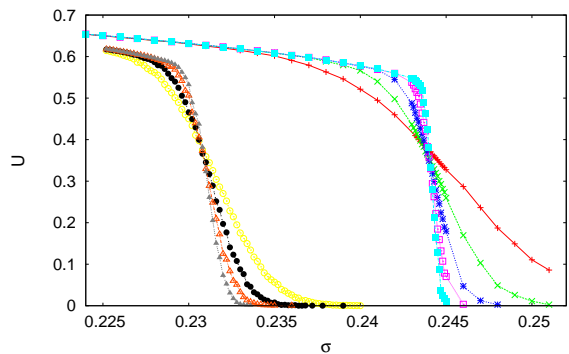}
   \caption{The fraction of surviving fibers ($U$) are plotted with load per fiber for $p=0.3$ both in 1d (left curves; $L=10000, 25000, 50000, 100000$)
            and 2d (right curves; $L=50, 100, 200, 400, 700$, $N=L\times L$).}
\label{compare.1d.2d.3}
\end{figure}
%%%%%%%%%%%%%%%%%%%%%%%%%%%%%%%%%%%%%%%%%%%%%%%%%%%%%%%%%%%%%%%%%%%%%%%%%%%%%%%%%%%%%%%%%%%%%%%%%%%%%%%%%%%%%%%%%%%%%%%%%%%%%%%%%%%%%%%%%%
%%%%%%%%%%%%%%%%%%%%%%%%%%%%%%%%%%%%%%%%%%%%%%%%%%%%%%%%%%%%%%%%%%%%%%%%%%%%%%%%%%%%%%%%%%%%%%%%%%%%%%%%%%%%%%%%%%%%%%%%%%%%%%%%%%%%%%%%%
\begin{figure}[tb]
\centering \includegraphics[width=8cm]{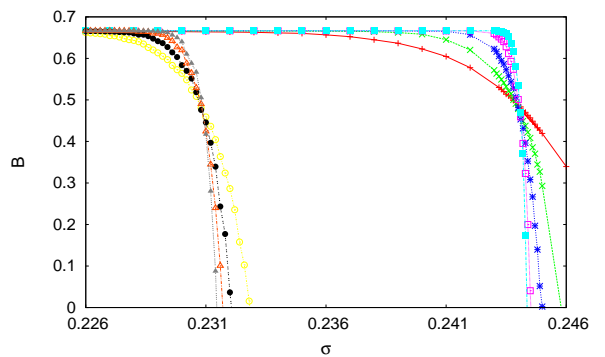}
   \caption{The Binder cumulants ($B$) are plotted with load per fiber for $p=0.3$ both in 1d (left curves; $L=10000, 25000, 50000, 100000$) 
            and 2d (right curves; $L=50, 100, 200, 400, 700$, $N=L\times L$).}
\label{compare.1d.2d.3.binder}
\end{figure}
%%%%%%%%%%%%%%%%%%%%%%%%%%%%%%%%%%%%%%%%%%%%%%%%%%%%%%%%%%%%%%%%%%%%%%%%%%%%%%%%%%%%%%%%%%%%%%%%%%%%%%%%%%%%%%%%%%%%%%%%%%%%%%%%%%%%%%%%%%
%%%%%%%%%%%%%%%%%%%%%%%%%%%%%%%%%%%%%%%%%%%%%%%%%%%%%%%%%%%%%%%%%%%%%%%%%%%%%%%%%%%%%%%%%%%%%%%%%%%%%%%%%%%%%%%%%%%%%%%%%%%%%%%%%%%%%%%%%
%%%%%%%%%%%%%%%%%%%%%%%%%%%%%%%%%%%%%%%%%%%%%%%%%%%%%%%%%%%%%%%%%%%%%%%%%%%%%%%%%%%%%%%%%%%%%%%%%%%%%%%%%%%%%%%%%%%%%%%%%%%%%%%%%%%%%%%%%
\begin{figure}[tb]
\centering \includegraphics[width=9cm]{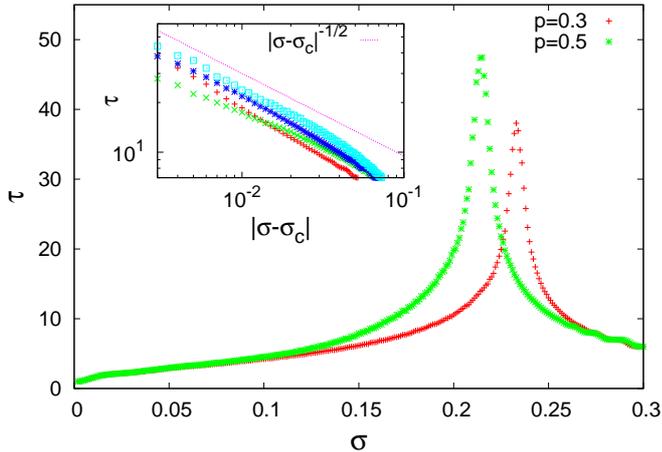}
   \caption{Relaxation time ($\tau$) is plotted for one dimensional heterogeneous model with $p=0.3, 0.5$. A power-law variation of the form 
            $\tau\sim |\sigma-\sigma_c|^{-z}$ was observed (for both side of the critical point), with $z=1/2$, similar to the global 
           load sharing model.}
\label{relax-1d}
\end{figure}
%%%%%%%%%%%%%%%%%%%%%%%%%%%%%%%%%%%%%%%%%%%%%%%%%%%%%%%%%%%%%%%%%%%%%%%%%%%%%%%%%%%%%%%%%%%%%%%%%%%%%%%%%%%%%%%%%%%%%%%%%%%%%%%%%%%%%%%%%%
%%%%%%%%%%%%%%%%%%%%%%%%%%%%%%%%%%%%%%%%%%%%%%%%%%%%%%%%%%%%%%%%%%%%%%%%%%%%%%%%%%%%%%%%%%%%%%%%%%%%%%%%%%%%%%%%%%%%%%%%%%%%%%%%%%%%%%%%%

While we plot the fraction ($U$) of surviving fibers with external load, to accurately determine the critical point
we also study the reduced fourth order Binder cumulant \cite{binder}
\begin{equation}
B=1-\frac{\langle U^4\rangle}{3\langle U^2\rangle^2},
\end{equation}
which has the property that its value is independent of system size at the critical point. This helps in
determining the critical point very accurately. In Fig. \ref{compare.1d.2d.3} we plot the fraction of surviving fibers with external
load and for different system sizes (at $p=0.3$ and $p=0.0$). Then in Fig. \ref{compare.1d.2d.3.binder} we plot the Binder cumulant for different system
sizes (also for $p=0.3$ and $p=0.0$). The common crossing point, which signifies size independent value, is the critical point. In this way
the phase boundary was determined.  

To determine the critical behavior,  we  plot the relaxation
time in Fig. \ref{relax-1d}. The relaxation time is basically the interval between two successive load increments. In the steady state when the external load is increased, the whole lattice is scanned once and the fibers having breaking threshold below the present load per fiber are broken. The load redistribution, however, is made after one full scan of the lattice, or in other words, parallel dynamics is followed. Now, one single scan of the entire lattice is taken as one Monte Carlo time step. In this way the relaxation time is estimated as the time steps required by the system to reach a steady state after a load increment. The relaxation time is expected to diverge near the critical point as $\tau\sim (\sigma_c-\sigma)^{-z}$,
From the simulations we see that $z=0.50\pm 0.01$, which is  in agreement with GLS exact result $z=1/2$. 
We have checked that these exponent values remain unchanged upto $p=0.95$ and beyond that no power law variation is
observed. Of course $p=0.95$ is the crossover point in the phase boundary, beyond which the system fails as soon as
any finite load is applied and there is no common crossing point of the Binder cumulant. This indicates that the 
mean field critical behavior is present upto the crossover point and beyond that it goes over to local load sharing behavior,
where of course critical behavior is not present. We also note that the crossover point is almost upto the point of
percolation. In one dimension, if the disorder has to percolate, its concentration will have to be unity. In this case
that value is almost reached.  
\subsection{Results in two dimensions}
\begin{figure}[tb]
\centering \includegraphics[width=9cm]{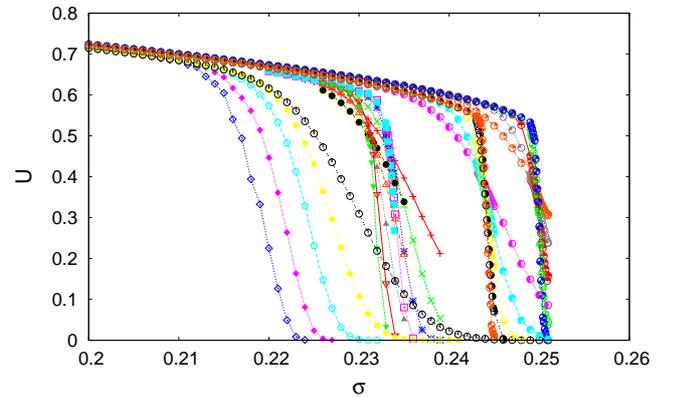}
   \caption{The fraction of surviving fibers are plotted against load per fiber for (left to right) $p=0.0, 0.3, 0.5, 0.52, 0.6$
            and for $L=50, 100, 200, 400, 700$ for each $p$ in two dimensions. Where the transition point became sharper with system size 
            for $p<p_c$, for $p>p_c$ the critical load goes to zero, signifying a crossover to LLS behavior.}
\label{compare.few}
\end{figure}

\begin{figure}[tb]
\centering \includegraphics[width=9cm]{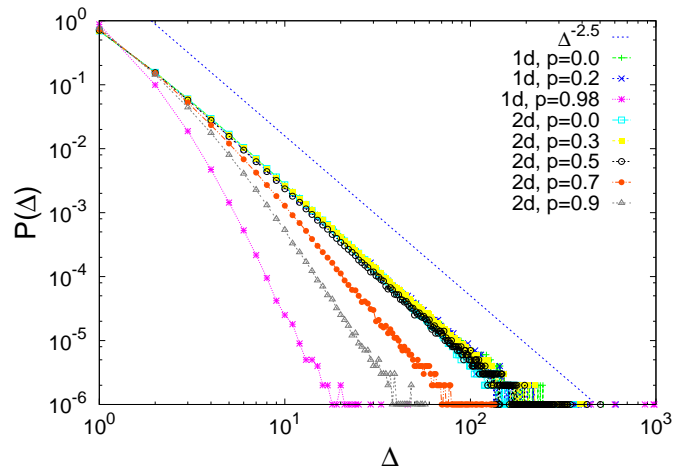}
   \caption{The avalanche size distribution is plotted for $p=0.0, 0.3, ,0.5, 0.7, 0.9$ for 2d and $p=0.0, 0.2, 0.98$ for 1d. It is seen that for $p<p_c$ the size distribution
follow the same power law with exponent $2.50\pm 0.01$ and for $p>p_c$, the distribution deviates from the scale free behavior. In simulations, load is increased upto the point when the weakest fiber breaks and number of broken fibers are counted until the system comes to a steady state.}
\label{avalanche}
\end{figure}

\begin{figure}[tb]
\centering \includegraphics[width=9cm]{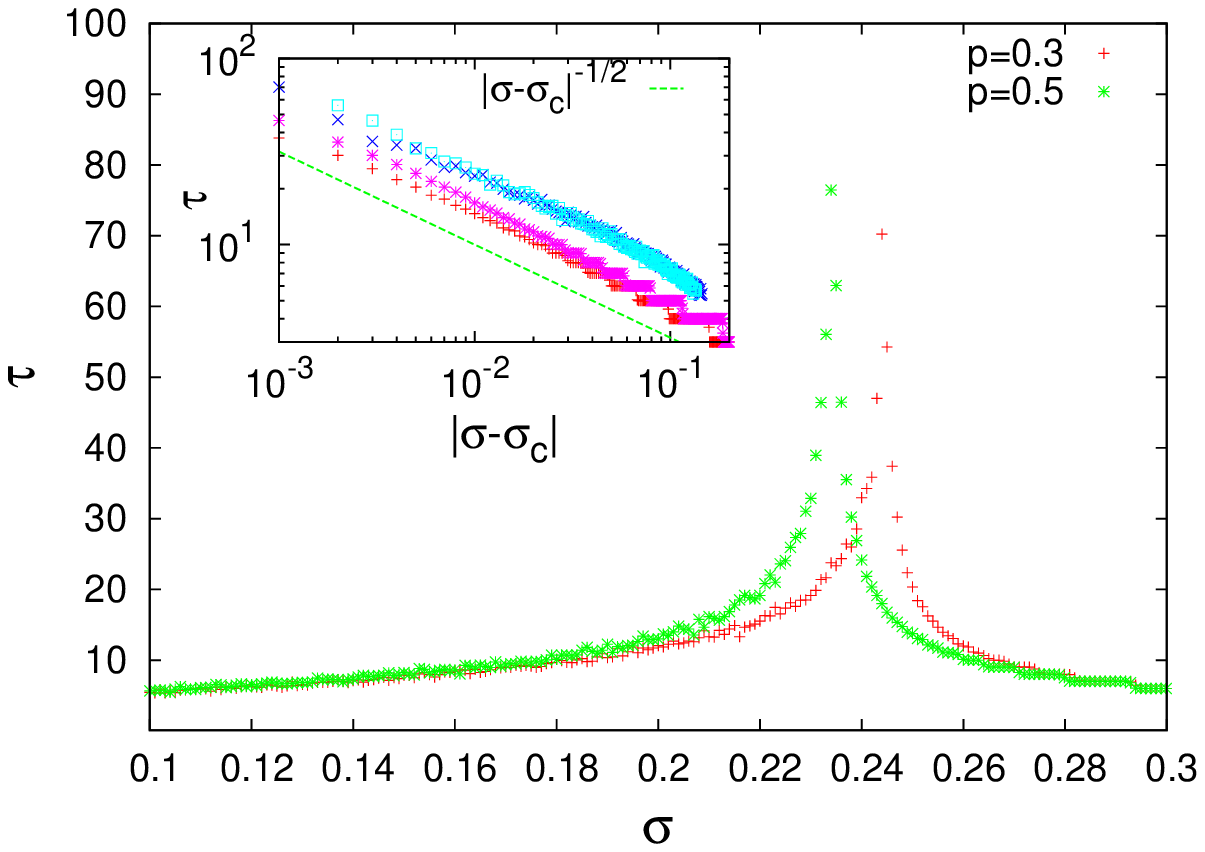}
   \caption{Relaxation time ($\tau$) is plotted for two dimensional heterogeneous model with $p=0.3, 0.5$. A power-law variation of the form 
            $\tau\sim |\sigma-\sigma_c|^{-z}$ was observed (for both side of the critical point), with $z=1/2$, similar to the global 
           load sharing model.}
\label{relax-2d}
\end{figure}

\begin{figure}[tb]
\centering \includegraphics[width=9cm]{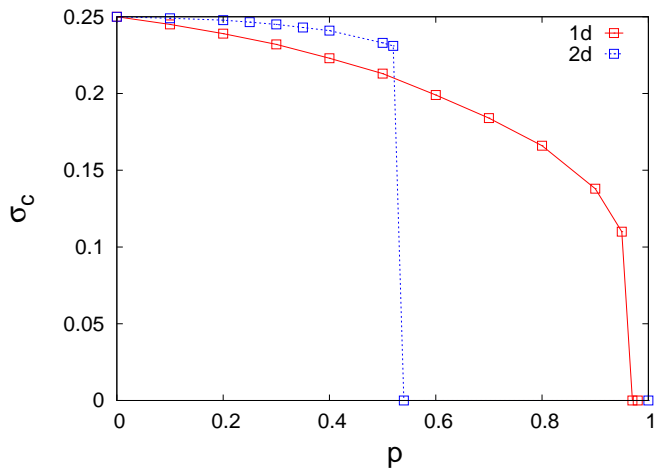}
   \caption{The phase diagrams for heterogeneous load sharing in one and two dimensions. The crossover points are at $p_c=0.95\pm0.01$
            and $p_c=0.53\pm 0.01$ respectively.}
\label{ph-dia-1d-2d}
\end{figure}

\noindent Similar to the case in one dimension, this heterogeneous load sharing mechanism can be followed in the more realistic 
case of two
dimensions as well. In this case, after each scan of the lattice, the clusters of the broken fibers are identified.
Then the fraction of the load to be shared locally is distributed along the cluster boundary and the rest is redistributed 
globally. As before, the load of an isolated fiber with local load sharing property will go to the surrounding cluster boundary of the patch of the broken fiber in which the isolated fiber will belong after its breaking. The fraction of surviving fibers with external load for various $p$ values are shown in Fig. \ref{compare.few}
%%%%%%%%%%%%%%%%%%%%%%%%%%%%%%%%%%%%%%%%%%%%%%%%%%%%%%%%%%%%%%%%%%%%%%%%%%%%%%%%%%%%%%%%%%%%%%%%%%%%%%%%%%%%%%%%%%%%%%%%%%%%%%%%%%%%%%%%%
%%%%%%%%%%%%%%%%%%%%%%%%%%%%%%%%%%%%%%%%%%%%%%%%%%%%%%%%%%%%%%%%%%%%%%%%%%%%%%%%%%%%%%%%%%%%%%%%%%%%%%%%%%%%%%%%%%%%%%%%%%%%%%%%%%%%%%%%%%
%%%%%%%%%%%%%%%%%%%%%%%%%%%%%%%%%%%%%%%%%%%%%%%%%%%%%%%%%%%%%%%%%%%%%%%%%%%%%%%%%%%%%%%%%%%%%%%%%%%%%%%%%%%%%%%%%%%%%%%%%%%%%%%%%%%%%%%%%
%%%%%%%%%%%%%%%%%%%%%%%%%%%%%%%%%%%%%%%%%%%%%%%%%%%%%%%%%%%%%%%%%%%%%%%%%%%%%%%%%%%%%%%%%%%%%%%%%%%%%%%%%%%%%%%%%%%%%%%%%%%%%%%%%%%%%%%%%
\begin{figure}[tb]
\centering \includegraphics[width=9cm]{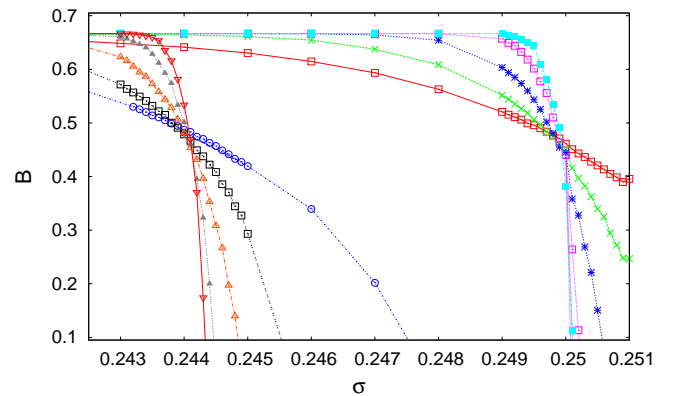}
   \caption{The Binder cumulants are plotted with load per fiber for $p=0.0, 0.3$ and different system sizes  ($L=50, 100, 200, 400, 700$)
             in two dimensions. 
             The clear common crossing point gives $\sigma_c$ in each case.}
\label{compare.0.3.binder}
\end{figure}
%%%%%%%%%%%%%%%%%%%%%%%%%%%%%%%%%%%%%%%%%%%%%%%%%%%%%%%%%%%%%%%%%%%%%%%%%%%%%%%%%%%%%%%%%%%%%%%%%%%%%%%%%%%%%%%%%%%%%%%%%%%%%%%%%%%%%%%%%%
%%%%%%%%%%%%%%%%%%%%%%%%%%%%%%%%%%%%%%%%%%%%%%%%%%%%%%%%%%%%%%%%%%%%%%%%%%%%%%%%%%%%%%%%%%%%%%%%%%%%%%%%%%%%%%%%%%%%%%%%%%%%%%%%%%%%%%%%%
%%%%%%%%%%%%%%%%%%%%%%%%%%%%%%%%%%%%%%%%%%%%%%%%%%%%%%%%%%%%%%%%%%%%%%%%%%%%%%%%%%%%%%%%%%%%%%%%%%%%%%%%%%%%%%%%%%%%%%%%%%%%%%%%%%%%%%%%%
\begin{figure}[tb]
\centering \includegraphics[width=9cm]{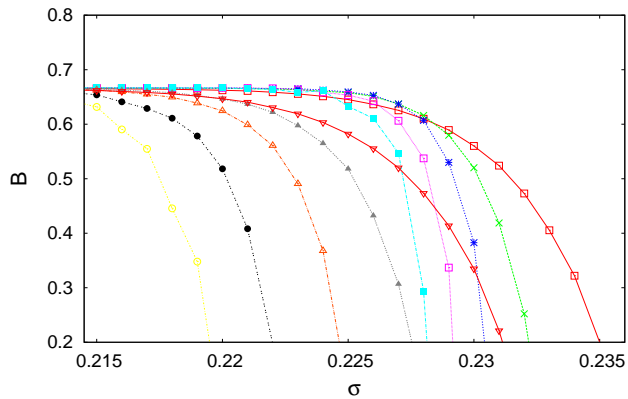}
   \caption{The Binder cumulants are plotted with load per fiber for $p=0.55, 0.60$ and different system sizes in two dimensions.
            In sharp contrast with the case of $p<p_c$, here we don't find any common crossing point for different sizes.}
\label{compare.55.6.binder}
\end{figure}
%%%%%%%%%%%%%%%%%%%%%%%%%%%%%%%%%%%%%%%%%%%%%%%%%%%%%%%%%%%%%%%%%%%%%%%%%%%%%%%%%%%%%%%%%%%%%%%%%%%%%%%%%%%%%%%%%%%%%%%%%%%%%%%%%%%%%%%%%%
%%%%%%%%%%%%%%%%%%%%%%%%%%%%%%%%%%%%%%%%%%%%%%%%%%%%%%%%%%%%%%%%%%%%%%%%%%%%%%%%%%%%%%%%%%%%%%%%%%%%%%%%%%%%%%%%%%%%%%%%%%%%%%%%%%%%%%%%%
As before, one can determine the critical point by the common crossing point of the Binder cumulant for different
system sizes (see Fig. \ref{compare.0.3.binder}). The critical point decreases with increasing fraction of the local load sharing fibers. But above
a threshold value $p_c=0.53\pm0.01$, there is no common crossing point of the Binder cumulant (see Fig. \ref{compare.55.6.binder}). This
suggests a crossover to local load sharing. This point is also slightly below the site percolation threshold ($p_c^0=0.5927\dots$) \cite{stauffer}.
The critical behavior, determined from the avalanche size distribution (Fig. \ref{avalanche}) and divergence of relaxation time 
(Fig. \ref{relax-2d}), is same as that of GLS model. The phase diagrams for one and two dimensions for this model is shown in Fig. \ref{ph-dia-1d-2d}.

\begin{figure}[tb]
\centering \includegraphics[width=9cm]{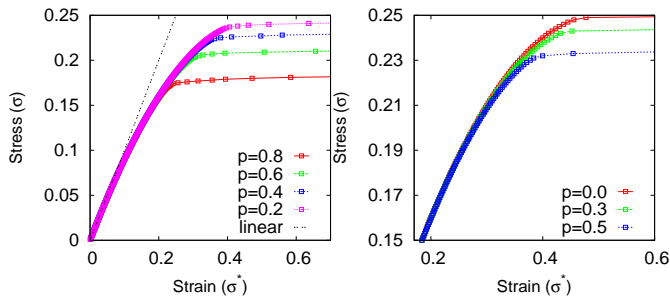}
   \caption{The left panel shows the stress ($\sigma$)-strain ($\sigma^*$) relation for different $p$ values for 1-d model. The linear relation
is plotted to show that the relation converges to linearity when $\sigma\to 0$. Similar behaviours are shown for 2-d version 
of the model on the right panel.}
\label{stress-strain}
\end{figure}

We have also shown the stress ($\sigma$)-strain ($\sigma^*$) (with $\sigma^*=\sigma/U$) relation for the model (see Fig. \ref{stress-strain}). While individual fibers behaves linearly, the collective stress-strain 
curve is non-linear but approaches linearity as $\sigma \to 0$. We show the behaviours for both one and two dimensions and for
different values of $p$, which is a measure of weakness of the material, and the breakdown occurs for lower stress as $p$ 
is increased. The non-linearity observed here is similar in form with that seen in Red Wildmoor sandstone \cite{earling} and this form seems to be better in describing the non-linearity than the one suggested in Ref. \cite{earling}.
%%%%%%%%%%%%%%%%%%%%%%%%%%%%%%%%%%%%%%%%%%%%%%%%%%%%%%%%%%%%%%%%%%%%%%%%%%%%%%%%%%%%%%%%%%%%%%%%%%%%%%%%%%%%%%%%%%%%%%%%%%%%%%%%%%%%%%%%%
%%%%%%%%%%%%%%%%%%%%%%%%%%%%%%%%%%%%%%%%%%%%%%%%%%%%%%%%%%%%%%%%%%%%%%%%%%%%%%%%%%%%%%%%%%%%%%%%%%%%%%%%%%%%%%%%%%%%%%%%%%%%%%%%%%%%%%%%%%
%%%%%%%%%%%%%%%%%%%%%%%%%%%%%%%%%%%%%%%%%%%%%%%%%%%%%%%%%%%%%%%%%%%%%%%%%%%%%%%%%%%%%%%%%%%%%%%%%%%%%%%%%%%%%%%%%%%%%%%%%%%%%%%%%%%%%%%%%
%%%%%%%%%%%%%%%%%%%%%%%%%%%%%%%%%%%%%%%%%%%%%%%%%%%%%%%%%%%%%%%%%%%%%%%%%%%%%%%%%%%%%%%%%%%%%%%%%%%%%%%%%%%%%%%%%%%%%%%%%%%%%%%%%%%%%%%%%
%%%%%%%%%%%%%%%%%%%%%%%%%%%%%%%%%%%%%%%%%%%%%%%%%%%%%%%%%%%%%%%%%%%%%%%%%%%%%%%%%%%%%%%%%%%%%%%%%%%%%%%%%%%%%%%%%%%%%%%%%%%%%%%%%%%%%%%%%%
%%%%%%%%%%%%%%%%%%%%%%%%%%%%%%%%%%%%%%%%%%%%%%%%%%%%%%%%%%%%%%%%%%%%%%%%%%%%%%%%%%%%%%%%%%%%%%%%%%%%%%%%%%%%%%%%%%%%%%%%%%%%%%%%%%%%%%%%%
%%%%%%%%%%%%%%%%%%%%%%%%%%%%%%%%%%%%%%%%%%%%%%%%%%%%%%%%%%%%%%%%%%%%%%%%%%%%%%%%%%%%%%%%%%%%%%%%%%%%%%%%%%%%%%%%%%%%%%%%%%%%%%%%%%%%%%%%%
%%%%%%%%%%%%%%%%%%%%%%%%%%%%%%%%%%%%%%%%%%%%%%%%%%%%%%%%%%%%%%%%%%%%%%%%%%%%%%%%%%%%%%%%%%%%%%%%%%%%%%%%%%%%%%%%%%%%%%%%%%%%%%%%%%%%%%%%%%
%%%%%%%%%%%%%%%%%%%%%%%%%%%%%%%%%%%%%%%%%%%%%%%%%%%%%%%%%%%%%%%%%%%%%%%%%%%%%%%%%%%%%%%%%%%%%%%%%%%%%%%%%%%%%%%%%%%%%%%%%%%%%%%%%%%%%%%%%
%%%%%%%%%%%%%%%%%%%%%%%%%%%%%%%%%%%%%%%%%%%%%%%%%%%%%%%%%%%%%%%%%%%%%%%%%%%%%%%%%%%%%%%%%%%%%%%%%%%%%%%%%%%%%%%%%%%%%%%%%%%%%%%%%%%%%%%%%
%%%%%%%%%%%%%%%%%%%%%%%%%%%%%%%%%%%%%%%%%%%%%%%%%%%%%%%%%%%%%%%%%%%%%%%%%%%%%%%%%%%%%%%%%%%%%%%%%%%%%%%%%%%%%%%%%%%%%%%%%%%%%%%%%%%%%%%%%%
\section{Dependence on loading process}
\noindent As mentioned before, there can be several ways in which external load can be applied in the system. The two
extreme cases are (i) increasing the load upto the point when the weakest fiber breaks and (ii) applying the desired
load directly on the intact fiber. Between these two ways, one can also adapt a third possibility when the load is
increased by equal amount in every step (after an avalanche stops). In GLS scheme, the fraction of surviving fibers
and for that matter the critical point would not depend on how one increases the load. However, the situation can change 
when there is local load sharing mechanism present (even partially). 
%%%%%%%%%%%%%%%%%%%%%%%%%%%%%%%%%%%%%%%%%%%%%%%%%%%%%%%%%%%%%%%%%%%%%%%%%%%%%%%%%%%%%%%%%%%%%%%%%%%%%%%%%%%%%%%%%%%%%%%%%%%%%%%%%%%%%%%%%
\begin{figure}[tb]
\centering \includegraphics[width=9cm]{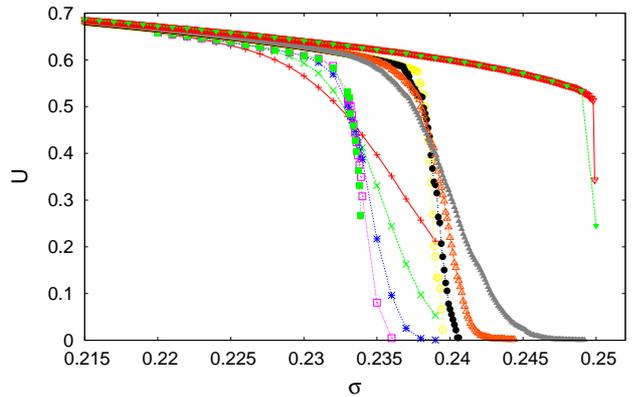}
   \caption{The fraction of surviving fibers are plotted against external load for two loading processes: equal load
              increment with $\delta=0.0001$ and direct loading, for $p=0.5$ and $0.0$. For $p=0.0$, the two curves
           for two loading processes exactly fall on top of each other, while those for $p=0.5$ are significantly different.}
\label{compare.5.5.0.0.diffload}
\end{figure}
%%%%%%%%%%%%%%%%%%%%%%%%%%%%%%%%%%%%%%%%%%%%%%%%%%%%%%%%%%%%%%%%%%%%%%%%%%%%%%%%%%%%%%%%%%%%%%%%%%%%%%%%%%%%%%%%%%%%%%%%%%%%%%%%%%%%%%%%%%
%%%%%%%%%%%%%%%%%%%%%%%%%%%%%%%%%%%%%%%%%%%%%%%%%%%%%%%%%%%%%%%%%%%%%%%%%%%%%%%%%%%%%%%%%%%%%%%%%%%%%%%%%%%%%%%%%%%%%%%%%%%%%%%%%%%%%%%%%

In the case of our model, we have studied the three ways of increasing the load and found that the critical point differs
for finite $p$ values. In Fig. \ref{compare.5.5.0.0.diffload} , we have plotted the fraction of surviving fibers  for equal load
increment and direct loading for $p=0.5$. The results show that the critical point is significantly different for the
two cases. We have also plotted the same cases for $p=0$ (GLS model) but it is clear that the load increment mechanisms
have no effect on the GLS version. This is probably due to the fact that spatial correlations in avalanches are
present in the local load sharing, which depends on the load increment method. But for GLS there is no spatial structure
as such. 
%%%%%%%%%%%%%%%%%%%%%%%%%%%%%%%%%%%%%%%%%%%%%%%%%%%%%%%%%%%%%%%%%%%%%%%%%%%%%%%%%%%%%%%%%%%%%%%%%%%%%%%%%%%%%%%%%%%%%%%%%%%%%%%%%%%%%%%%%
\begin{figure}[tb]
\centering \includegraphics[width=9cm]{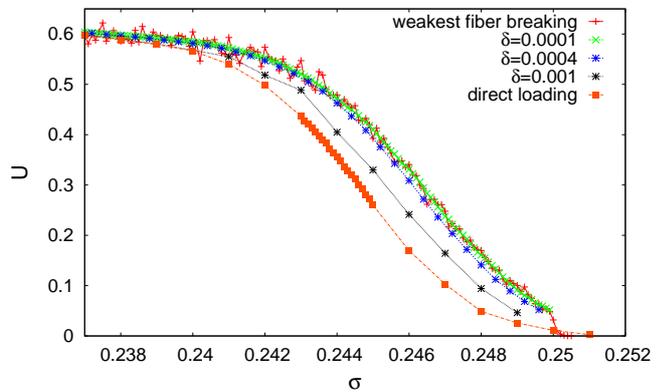}
   \caption{The fraction of surviving fibers are plotted against external load for different loading processes when $p=0.3$.
            For equal load increment with $\delta=0.0001$ almost matches with the curve obtained for loading upto weakest fiber breaking.
           The equal step load increment with $\delta=0.0004, 0.001$ are also shown. They are in between the $\delta=0.0001$ curve
           and that due to direct loading.}
\label{l100.compare}
\end{figure}
%%%%%%%%%%%%%%%%%%%%%%%%%%%%%%%%%%%%%%%%%%%%%%%%%%%%%%%%%%%%%%%%%%%%%%%%%%%%%%%%%%%%%%%%%%%%%%%%%%%%%%%%%%%%%%%%%%%%%%%%%%%%%%%%%%%%%%%%%%
%%%%%%%%%%%%%%%%%%%%%%%%%%%%%%%%%%%%%%%%%%%%%%%%%%%%%%%%%%%%%%%%%%%%%%%%%%%%%%%%%%%%%%%%%%%%%%%%%%%%%%%%%%%%%%%%%%%%%%%%%%%%%%%%%%%%%%%%%

In Fig. \ref{l100.compare} we plot, for a given system size and $p$ value, the fraction of surviving fibers for different loading methods.
We find that the weakest fiber breaking and the direct loading are two limits, and equal load increment method interpolates
between the two. For very small value of the fixed increment rate ($\delta=0.0001$), the curve is almost identical with
that in weakest fiber breaking. This is because, for $\delta\sim 1/N$, the equal load increment method should be effectively
same as that of weakest fiber breaking. 

Finally, we note that although the critical point is dependent on the load increment mechanisms for finite $p$ values,
the crossover point and the critical behavior below the crossover point is unaffected by the load sharing mechanisms. 
The results presented in the previous section are for direct loading, except for the avalanche size distribution, which
is for weakest fiber breaking.

%%%%%%%%%%%%%%%%%%%%%%%%%%%%%%%%%%%%%%%%%%%%%%%%%%%%%%%%%%%%%%%%%%%%%%%%%%%%%%%%%%%%%%%%%%%%%%%%%%%%%%%%%%%%%%%%%%%%%%%%%%%%%%%%%%%%%%%%%
\begin{figure}[tb]
\centering \includegraphics[width=9cm]{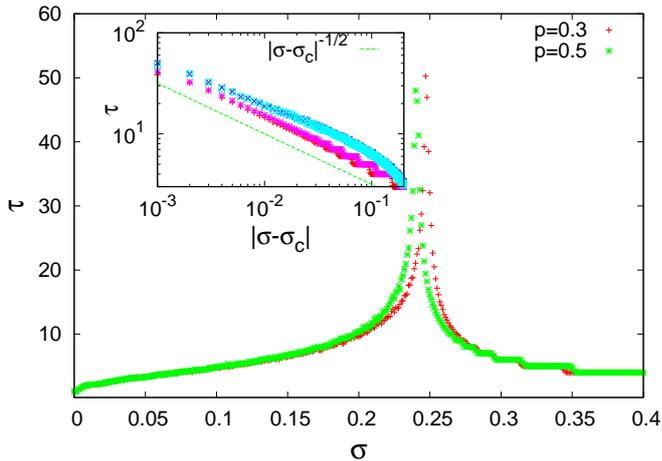}
   \caption{Relaxation time ($\tau$) is plotted for two dimensional homogeneous model \cite{inter2} with $p=0.3, 0.5$. A power-law variation of the form 
            $\tau\sim |\sigma-\sigma_c|^{-z}$ was observed (for both side of the critical point), with $z=1/2$, similar to the global 
           load sharing model.}
\label{relax-2d-g}
\end{figure}
%%%%%%%%%%%%%%%%%%%%%%%%%%%%%%%%%%%%%%%%%%%%%%%%%%%%%%%%%%%%%%%%%%%%%%%%%%%%%%%%%%%%%%%%%%%%%%%%%%%%%%%%%%%%%%%%%%%%%%%%%%%%%%%%%%%%%%%%%%
%%%%%%%%%%%%%%%%%%%%%%%%%%%%%%%%%%%%%%%%%%%%%%%%%%%%%%%%%%%%%%%%%%%%%%%%%%%%%%%%%%%%%%%%%%%%%%%%%%%%%%%%%%%%%%%%%%%%%%%%%%%%%%%%%%%%%%%%%
\section{Comparison with mixed load sharing model}
\noindent The present model (refer to this as `$p$' model) is also an interpolation scheme between LLS and GLS. 
In fact this is the heterogeneous version 
of the mixed mode load sharing (MMLS) model \cite{inter2}, where after failure of every fiber, a fraction $g$ of the load of a broken fiber 
is distributed locally and the
rest is distributed globally (we refer to this as `g' model). This model was studied in one dimension in Ref. \cite{inter2}. 
We have studied it in two dimensions as well. Although the amount of load shared locally on average is same for the 
two models, the local fluctuations
introduced in the present version makes it significantly different from the earlier one. 

The relaxation time divergence of this model in two dimensions (Fig. \ref{relax-2d-g}), shows the same critical behavior
as that of the one dimensional version (and GLS model).
In Fig. \ref{1d-2d-g-p} we compare the phase diagrams of the two models in one and two dimension, which shows the effect of 
heterogeneity as the phase boundaries do not coincide. The phase boundary and critical behavior 
of MMLS model in one dimension was reported in ref. \cite{inter2}. Here we determine those in two dimensions.  

\section{Summary and Discussions}
\noindent There are two extreme cases of load sharing (GLS and LLS) in the fiber bundle models. Since load sharing
mechanism is the only source of cooperativity in these models, in that sense the load sharing mechanisms 
determine the spatial fluctuations. Apart from the theoretical significances of the possible
spatial correlations, the load sharing mechanism is of vital practical importance, since it determines the strength 
of the material (irrespective of threshold distribution). 

While the completely global load sharing is idealised case, the completely local load sharing has no finite critical
point. Hence, there have been several interpolation mechanisms to match realistic situations \cite{inter1,inter2,inter3,inter4}. 
However, in
all such cases the load sharing methods is same throughout the sample. But here we argue that local disorders in the
materials can make the load sharing spatially heterogeneous. In this regards, we have proposed a simple heterogeneous
load sharing model that gives crossover behavior at a non-trivial concentration value of the disorder. 
%%%%%%%%%%%%%%%%%%%%%%%%%%%%%%%%%%%%%%%%%%%%%%%%%%%%%%%%%%%%%%%%%%%%%%%%%%%%%%%%%%%%%%%%%%%%%%%%%%%%%%%%%%%%%%%%%%%%%%%%%%%%%%%%%%%%%%%%%
\begin{figure}[tb]
\centering \includegraphics[width=9.2cm]{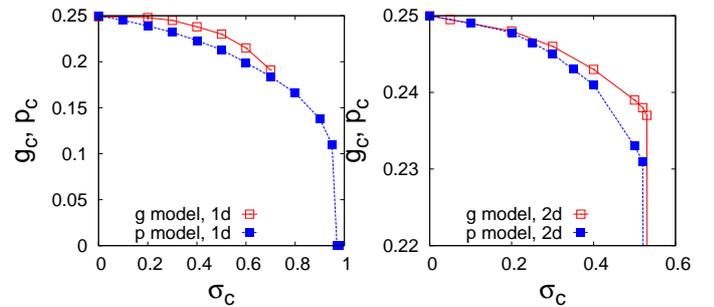}
   \caption{The left panel shows the comparisons of the phase diagram of the heterogeneous and the homogeneous models in one dimension and the right panel shows the comparisons of the phase diagram of the heterogeneous and the homogeneous models in two dimension. The $p_c - \sigma_c$ corresponds to the critical values of $\sigma$ for the given $p$ in the models introduced here, while the $g_c - \sigma_c$ line corresponds to the same for different values of $g$ in MMLS model \cite{inter2}.}
\label{1d-2d-g-p}
\end{figure}
%%%%%%%%%%%%%%%%%%%%%%%%%%%%%%%%%%%%%%%%%%%%%%%%%%%%%%%%%%%%%%%%%%%%%%%%%%%%%%%%%%%%%%%%%%%%%%%%%%%%%%%%%%%%%%%%%%%%%%%%%%%%%%%%%%%%%%%%%%
%%%%%%%%%%%%%%%%%%%%%%%%%%%%%%%%%%%%%%%%%%%%%%%%%%%%%%%%%%%%%%%%%%%%%%%%%%%%%%%%%%%%%%%%%%%%%%%%%%%%%%%%%%%%%%%%%%%%%%%%%%%%%%%%%%%%%%%%%

We have studied this model in both one and two dimensions. In one dimension, we have found that the critical disorder
concentration at which the crossover occurs is $p_c=0.95\pm 0.01$, which is very close to the trivial percolation threshold
in one dimension (unity). The phase diagram was determined using Binder cumulant. It is seen that the critical behavior,
in terms of avalanche size distribution and relaxation time divergence, remains that of the GLS model upto the crossover
point (see Fig. \ref{relax-1d}). The same behavior is seen in two dimensions. However, the crossover point is now $p_c=0.53\pm0.01$, which is
again close (and slightly below) the site percolation threshold in square lattice ($0.5927\dots$) \cite{stauffer}. The fact that the 
crossover occurs even before the percolation may be due to the fact that even for local load sharing, the load transfer 
may some times be long ranged due to the absence of neighbors. Of course the percolation threshold should be the upper bound 
for the crossover, since beyond the percolation threshold the local load sharing fibers will form a compact cluster, the size
of which will scale as the system size and therefore, that cluster will fail at any finite load due to the same reason as a
fully local load sharing system always fails.

We have also noted that a finite local load sharing fraction would lead to dependence on loading history. For a GLS
model, the fraction of surviving fibers and the critical point was independent of loading history. But as soon as
a local load sharing fraction is introduced, these quantities become loading history dependent (see Fig. \ref{l100.compare}).
 This is an
important factor, because in a real material it is not possible to determine the microscopic load sharing mechanism
directly. But this is a macroscopic manifestation of the microscopic load sharing process. And for the strength of the 
material, it is very important whether the load sharing is global or local. Because, as we see from the phase diagram 
(see Fig. \ref{ph-dia-1d-2d}), the system breaks down as soon as any load is applied much before the load sharing is completely local. 
It is in fact close to (below) the percolation threshold in the respective lattices. 

Finally, we have compared our model with another interpolation scheme, which is in fact the homogeneous version of the 
present model. Although the average load sharing of local and global fractions are same, due to local fluctuations
in the present model, the phase boundaries are different (see Fig. \ref{1d-2d-g-p}). 

In summary, we have proposed a heterogeneous load sharing mechanism for fiber bundle models to account for the
local disorder concentration in the bottom plate of the fiber bundle models. 
For finite fraction of local load sharing, the model has loading history dependent response.
We have found numerically, that
 the critical behavior crosses over from GLS (mean field) to LLS at some effective site percolation threshold. 
The critical behavior across the phase boundary (Fig. \ref{1d-2d-g-p}) is of mean field type (relaxation time divergence
exponent $z=0.50\pm 0.01$ and avalanche size distribution has an exponent $2.50\pm 0.01$).

\vskip0.5cm
\thanks
\noindent We acknowledge Srutarshi Pradhan for discussions and suggestions and Erling Fj{\ae}r for pointing out Ref. \cite{earling}. 
The work was finalized during both
the authors' visit to SINTEF, Norway. The authors also acknowledge participants of the INDNOR Project
meeting for discussions and BKC's JC Bose Fellowship (DST, Govt. of India) for partial support.

%
%
% BibTeX users please use
% \bibliographystyle{}
% \bibliography{}

\end{document}